\documentclass[aps,twocolumn,showpacs]{revtex4}% Physical Review B
\usepackage{graphicx}% Include figure files
\begin{document}
\title{Resonant tunnelling between Luttinger liquids: solvable case}
\author{A. Komnik$^{1}$ and A. O. Gogolin$^{2}$}

\affiliation{$^{1}$Physikalisches Institut, Albert--Ludwigs--Universit\"at,
D--79104 Freiburg, Germany \\
$^{2}$Department of Mathematics, Imperial College, 180 Queen's Gate,
London SW7 2BZ, United Kingdom}

\date{\today}

\begin{abstract}

We discuss the conductance of a Luttinger liquid interrupted
by a quantum dot containing a single resonant level.
Using bosonisation and re-fermionisation methods, we
find a mapping to a Kondo-type problem which possesses 
a non-trivial Toulouse-type solvable point.
At this point, we obtain an analytic expression for 
the non-linear current-voltage characteristics and analyse 
the differential conductance and the width of the resonance peak  
as functions of bias and gate voltages, temperature, and barrier asymmetry. 
We also determine the exact scaling function for the linear conductance.
\end{abstract}

\maketitle

The field of one-dimensional interacting metallic systems recently 
experienced another revival as single--wall carbon nanotubes (SWNTs)
have been found to display transport properties consistent with 
the Luttinger liquid (LL) theory \cite{dekker}. 
While the electrical transport through clean SWNTs has been
investigated in different independent experiments, 
the transport properties of SWNTs with impurities 
(or in more complicated set-ups) are still to be studied in detail. 
Progress was recently made in this direction: 
in \cite{postma} the manufacture of 
quantum dots on the nanotube basis was reported. 
Surprisingly, the authors found that the transport 
is dominated by a coherent transmission (or resonant tunnelling)
in a wide parameter range.

The presence of a resonant level is known to enhance conductance. 
Indeed, for non-interacting electrons the local level
hybridises with the conduction band causing a Lorentzian
shaped peak in the density of states (the conductance
being related to the Breit--Wigner scattering cross-section
via the Landauer formula).
Unless the system is exactly at resonance, the picture 
remains qualitatively the same for the
case of interacting electrons, even LLs \cite{kanefisher}. 
Therefore one expects the conductance to increase upon
lowering the temperature.
On the other hand, at low temperatures, the
conductance is known to vanish (unless exactly at resonance) 
due to the effective enhancement of backscattering processes 
specific for the LLs.
Hence a non-monotonic behaviour of the linear conductance
as a function of temperature. 
The limiting cases has been thoroughly studied in Refs. 
\cite{kanefisher,NG,furusakidot,nayak} but the 
full description of the cross-over remains an open
problem. 

Recently, Nazarov and Glazman (NG) calculated the cross-over
conductance in the weak electron-electron interaction limit
(when the LL parameter $g$ is close to 1) by using the
Landauer type approach supplemented by renormalisation group \cite{NG}.
In this paper, we wish to discuss the opposite limit of strong
interactions, which is relevant for such systems as SWNTs. 
We shall concentrate on the special value of the coupling $g=1/2$ and
present an explicit solution of the problem at this point.
At this particular value of the LL parameter 
and when the resonant level energy 
is tuned to match the equilibrium chemical potentials in the leads, 
the resonant tunnelling process is
marginally relevant and its amplitude
increases logarithmically upon lowering the energy
scale \cite{kanefisher,matveevdot}. 
It turns out that at low temperatures in the linear regime (i.e. in the
limit of small bias voltage) the sequential tunnelling dominates
the transport for $g<1/2$, while above that value the resonant transmission
wins over \cite{furusakidot,kanefisher}. 
Hence, apart from being supplementary to NG results, 
the exact solution at $g=1/2$ yields 
insights into the interplay between these two transport mechanisms. 

We model the system by a resonant level (which can also be regarded as a 
single state quantum dot, so we use both terms) 
coupled to interacting leads, which is described by the following Hamiltonian
(we ignore the spin degree of freedom throughout the paper),
\begin{eqnarray}                     \label{H0}
 H = H_K + H_t + H_C \, ,
\end{eqnarray}
where $H_K$ is the kinetic part, 
$H_K = \Delta d^\dag d + \sum_{i=R,L} H_0[\psi_i]$, 
describing the the electronic degrees of freedom in the leads $H_0[\psi_i]$,
and the resonant level with energy $\Delta$ with the corresponding electron 
operators being $d^\dag,d$. The dot can be populated from
either of the two leads ($i=R,L$) via electron tunnelling with amplitudes
$\gamma_i$, $H_t = \sum_i  \gamma_i[ d^\dag \psi_i(0) + \mbox{h.c.}]$.
Here $H_C$ describes the electrostatic 
Coulomb interaction between the leads and the dot, 
$H_C =  \lambda_C d^\dag d \, \sum_i \, \psi_i^\dag(0) \psi_i(0)$.
This interaction is a new ingredient we have introduced, absent in
\cite{kanefisher} and \cite{NG}. It does not, however, affect the
universality as we shall show. 
The contacting electrodes are supposed to be one-dimensional
half-infinite electron systems. We model them 
by chiral fermions living in an infinite system: the negative half-axis
then describes the particles moving towards the boundary, 
while the positive half-axis
carries electrons moving away from the end of the system. 
In the bosonic representation $H_0[\psi_i]$ are diagonal even in presence 
of interactions
(for a recent review see e.g. \cite{book};
we set the renormalised Fermi velocity $v=v_F/g=1$, 
the bare velocity being $v_F$): 
%\begin{eqnarray}                 \label{Hi}
 $H_0[\psi_i] = (4 \pi)^{-1} \int \, dx \, [\partial_x \phi_i(x)]^2$.
%\end{eqnarray} 
Here the phase fields $\phi_i(x)$ describe the slow varying
spatial component of the electron density (plasmons), 
$\psi^\dag_i(x) \psi_i(x) = \partial_x \phi_i(x)/2 \pi \sqrt{g}$.
The electron field operator at the boundary is given by\footnote{Strictly
speaking $\psi(x=0)=0$, so we 
assume that the tunnelling takes place at the second last site of the 
corresponding lattice model, at $x=\pm a_0$. Also, we ignore the Klein factors
as they can be absorbed into Eqs.(\ref{newfermions}) 
disappearing from the analysis.},
%\begin{eqnarray}                           \label{psioperator}
$\psi_i(0) = e^{i \phi_i(0)/\sqrt{g}}/\sqrt{2 \pi a_0}$,
%\end{eqnarray}
where $a_0$ is the lattice constant of the underlying lattice model. 
Here $g$ is the conventional LL parameter (coupling constant) 
\cite{book,kanefisher}. In the chiral formulation the bias voltage amounts 
to a difference in the densities of the incoming particles in both channels 
far away from the constriction \cite{eggergrabert}. The current is then 
proportional to the difference betwee the densities of 
incoming and outgoing particles within each channel.

To the best of our knowledge, Hamiltonian (\ref{H0}) cannot be solved exactly 
even in the $g=1$ case as long as $\lambda_C$ remains finite. However, after
a transformation of $d^\dag$ and $d$ operators to the spin representation 
of the form
%\begin{eqnarray}
% \left\{  \begin{array}{l}
                  $S_x = (d^\dag + d)/2$,
%                  \\
                  $S_y = - i (d^\dag - d)/2$,
%                  \\
                  $S_z = d^\dag d - 1/2$,
%                  \end{array} \right.
%           \; , \;
%\end{eqnarray}
one immediately observes that  the $\lambda_C$ term is 
analogous to the $S_z$--spin density coupling in the Kondo problem.
The latter is known to be explicitly solvable at a 
particular value of the longitudinal coupling: 
the Toulouse limit (see e.g. \cite{book}).  
Let us perform a similar calculation. 
As a first step we introduce new symmetric and antisymmetric 
fields
$\phi_\pm = (\phi_L \pm \phi_R)/\sqrt{2}$, which still fulfill the bosonic
commutation relations. Then we apply the transformation $H'=U^\dag H
U$ with $U=\exp( i S_z \phi_+/ \sqrt{2 g})$ \cite{emerykivelson},
which changes the kinetic and the Coulomb coupling parts of the full 
Hamiltonian to (we drop a constant contribution)  
\begin{eqnarray}
 H_K'+ H_C' = H_K + 
 (\lambda_C/\pi \sqrt{2g} - \sqrt{2/g}) S_z \partial_x
 \phi_+(0) \, , \nonumber 
\end{eqnarray}
and the tunnelling part (terms containing $\gamma_i$) to
\begin{eqnarray}
 H_t' &=& (2 \pi a_0)^{-1/2} \Big[ S_+ (\gamma_L e^{i \phi_-/\sqrt{2g}} +
 \gamma_R e^{-i \phi_-/\sqrt{2 g}}) 
 \nonumber \\
 &+& (\gamma_L e^{-i \phi_-/\sqrt{2g}} +
 \gamma_R e^{i \phi_-/\sqrt{2g}}) S_- \Big] \, ,
\end{eqnarray}
where $S_{\pm}= S_x \pm i S_y =d^\dag,d$.
At the point $g=1/2$ one can 
re-fermionise the problem by defining new operators
\begin{eqnarray}              \label{newfermions}
\psi_\pm = e^{i \phi_\pm}/\sqrt{2 \pi a_0} \, ,
\end{eqnarray}
which  fulfill standard fermionic commutation relations. With the help of
the particle density operator $\psi^\dag_\pm \psi_\pm =
\partial_x \phi_\pm/2 \pi$ we can immediately write down the refermionised
Hamiltonian, 
\begin{eqnarray}                   \label{Htransformed}
 H &=& H_0[\psi_\pm] + (\lambda_C - 2 \pi) 2S_z \psi_+^\dag \psi_+ + \Delta S_z
 \nonumber \\
 &+& S_+ (\gamma_L \psi_- + \gamma_R \psi^\dag_-) + (\gamma_L \psi^\dag_- +
 \gamma_R \psi_-) S_- \, .
\end{eqnarray}
In the case of the symmetric coupling $\gamma_L=\gamma_R$ this Hamiltonian 
is similar to that of the two-channel Kondo problem and, at the Toulouse point
$\lambda_C = 2 \pi$, can be solved exactly (out of equilibrium) 
using the method of Ref.\cite{SH}. 
The novel ingredient in the following analysis is the
extension to the asymmetric case. 
To take advantage of the Toulouse point we set the 
Coulomb coupling amplitude to $2\pi$ in what follows.
This not only removes the four fermion interaction
but decouples the `$\pm$' channels making the `$+$' channel free
(i.e. decoupled from the dot variables). 

As we already mentioned, 
due to the linear dispersion relation, 
the current through the system is proportional to the
difference between the densities of particles moving towards the dot 
and away from it in either of the channels. Due to the chiral geometry we 
then have 
 $I \sim \psi_L^\dag \psi_L^{\phantom{\dag}}(-\infty) 
- \psi_L^\dag \psi_L^{\phantom{\dag}}(\infty)$ , 
which, being transformed to `$\pm$' channels, results in
 $I \sim \psi_-^\dag \psi_-^{\phantom{\dag}}(-\infty) - 
\psi_-^\dag \psi_-^{\phantom{\dag}}(\infty)$. 
Since the `$+$' channel is free, it doesn't  contribute to the above formula. 
As the `$-$' channel is also free when away from the dot,
in order to calculate the current we only need to know the scattering matrix
of `$-$' fermions determined by Hamiltonian (\ref{Htransformed}). 
The chemical potential of the incoming particles is determined by the 
bias voltage. 
Hence, the current is given by (we measure voltage in energy units,
i.e. set $e=1$)
\begin{eqnarray}                          \label{IVdefinition}
 I(V) = G_0 \int \, d \omega \, T(\omega) [n_F(\omega-V) - n_F(\omega)] \, 
\end{eqnarray}
where $n_F$ denotes the Fermi distribution function and 
$D(\omega)=1-T(\omega)$ is the
energy dependent penetration coefficient of the `$-$' particles from $x<0$ to
$x>0$. The pre-factor $G_0 = e^2/h$ is fixed by the requirement that at zero
transmission $D(\omega)=0$ (or perfect transmission of the whole structure)
one obtains the correct conductance. 
%of a single free channel. 
\begin{figure}
\vspace*{0.5cm}
\includegraphics[scale=0.38]{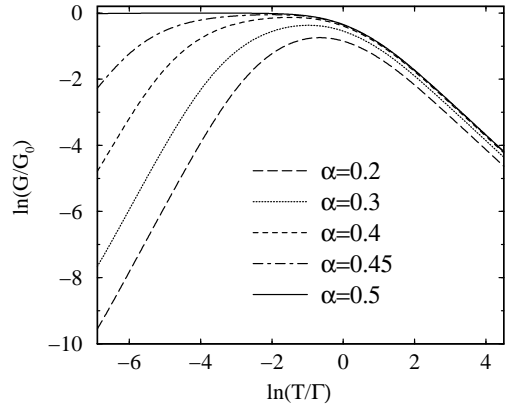}
\caption[]{\label{dcond} 
Linear differential conductance at the resonance $\Delta=0$ as a function 
of temperature for different values of the asymmetry parameter $\alpha$.}
\vspace*{-0.5cm}
\end{figure}

The easiest way to obtain the transmission coefficient is the equations of
motion method. Since we have two types of operators: for the 
electrons of the `$-$' channel and for the resonance level (we go back 
to the original $d^\dag, d$ operators), we need two equations of motion, 
\begin{eqnarray}                    \label{EoMs}
 i \partial_t \psi_-(x) &=& -i \partial_x \psi_-(x) + \delta(x)(\gamma_L d -
 \gamma_R d^\dag) \, , \nonumber \\
 i \partial_t d &=& \Delta d + \gamma_L \psi_-(0) + 
\gamma_R \psi_-^\dag(0) \, . 
\end{eqnarray}
Integrating the first one around $x=0$ we obtain 
%with respect to $x$ from $-\epsilon$
%to $\epsilon$ and then sending $\epsilon$ to zero we obtain
\begin{eqnarray}
 i [\psi_-(0^+)-\psi_-(0^-)] = \gamma_L d - \gamma_R d^\dag \, . 
\end{eqnarray}
%where $0^\pm$ denotes positive (negative) infinitesimal.
Acting with $\partial_t^2 + \Delta^2$ on both sides of this relation yields
\begin{eqnarray}                   \label{polnoye}
 &&(\partial_t^2 + \Delta^2)[\psi_-(0^+)-\psi_-(0^-)] \\ \nonumber
 &=& -
 [(\gamma_L^2+\gamma_R^2)\partial_t + i \Delta (\gamma_R^2-\gamma_L^2)]
 \psi_-(0) - 2 \gamma_R \gamma_L \partial_t \psi^\dag_-(0) \, .
\end{eqnarray}
Now we can insert into this relation the momentum decomposition of the field
operator $\psi_-$ 
\begin{eqnarray}                    \label{partialdecomposition}
 \psi_-(x,t) = \int \, \frac{d k}{2 \pi} e^{i k (t-x)} 
\left\{  \begin{array}{l}
                  a_k \, \, \, \mbox{for} \, x<0
                  \\
                  b_k \, \, \, \mbox{for} \, x>0
                  \end{array} \right. \, .
\end{eqnarray}
Because the dispersion relation is linear, $\omega = v k = k$, 
we can use $\omega$ as the momentum variable as well as the energy variable. 
Inserting Eq.(\ref{partialdecomposition}) into Eq.(\ref{polnoye})
and using $\psi_-(0) = [\psi_-(0^+) + \psi_-(0^-)]/2$ results in
\begin{eqnarray}                    \label{finalequation}
E(b_\omega - a_\omega) = -i \beta_+ (a_\omega + b_\omega) + i \gamma 
(a^\dag_{-\omega} + b^\dag_{-\omega})
\, , 
\end{eqnarray}
where we introduced the following objects: $E=\Delta^2-\omega^2$, 
$\beta_\pm = [(1-2\alpha)\Delta \pm \omega]/2$,  
$\gamma = \omega \sqrt{\alpha(1-\alpha)}$,
and $\alpha =\gamma_L^2/(\gamma_L^2 + \gamma_R^2)$ (the asymmetry parameter). 
From now on $\omega$, $\Delta$,
the bias voltage $V$, and the temperature $T$ are all measured in units of 
$\Gamma = \gamma_L^2 + \gamma_R^2$. Considering in addition to
Eq.(\ref{finalequation}) its complex conjugate for $-\omega$ we establish a 
relation
between the amplitudes of the incoming ($a_\omega$) and transmitted
($b_\omega$) particle fluxes. The
transmission coefficient can then be read off as follows:
\begin{eqnarray}                       \label{1-D}
 &&T(\omega) \\ \nonumber 
 &=&\frac{ 4 \gamma^2 E^2}{(E^2 + \beta_+^2)(E^2 + \beta_-^2) + 2
 \gamma^2 (E^2 + \beta_- \beta_+) + \gamma^4} \, ,
\end{eqnarray}
This equation, accompanied by Eq.(\ref{IVdefinition}), 
provides all informations about the transport properties
of the system and is the central result of this paper. 
The experimentally relevant quantity is the 
differential conductance $G=dI/dV$. 
At zero temperature, Eq.(\ref{IVdefinition}) considerably simplifies and,
differentiating with respect to the bias voltage, one immediately
finds that $G/G_0 = T(V)$.
In the case when the couplings between the dot and the leads are perfectly
symmetric and one of the chemical potentials matches $\Delta$, $G$ reaches
the maximal value of $G_0$. 
This is a typical signature of the resonant tunnelling effect 
usually encountered in transport phenomena in 
double-barrier structures \cite{datta}. 
In fact, our system \emph{is} a model for such a structure with 
one single state between the barriers. 

The interplay between the LL's enhancement of the 
backscattering at low temperatures 
(resulting in decreasing conductance) and the more standard 
Breit-Wigner physics  emerging in the resonant tunnelling can be 
seen in Fig.\ref{dcond}. 
As predicted in Refs.\cite{kanefisher,NG}, 
in the symmetric case $\alpha=0.5$ and for $\Delta=0$, 
the conductance saturates at low temperatures to its maximal value. 
In the presence of an asymmetry $G$ does not saturate any 
more and vanishes as a
power--law towards $T=0$ with the exponent $2$. 
This value is equal to twice the density of
states exponent $\nu$ of the LL with an open boundary:
$\nu=1/g-1$ \cite{book}, which in our case is equal to 1. 
This fact indicates that in this regime the electrons are transferred 
through the system in a single stage process \cite{NG}, 
so that the internal structure of the dot does not matter any more. 
Contrary to Eq.(9) of NG, the high 
temperature ($T\gg 1$) evolution of the conductance follows the law 
$G/G_0 \sim1/T$. 
The reason is that the problem maps onto a free-fermion
one, for which the $1/T$ behaviour is inevitable.
Note that, in the language of the original model, this
corresponds to tunnelling of \emph{composite} objects
from one LL into the other.  

\begin{figure}
\vspace*{0.5cm}
\includegraphics[scale=0.38]{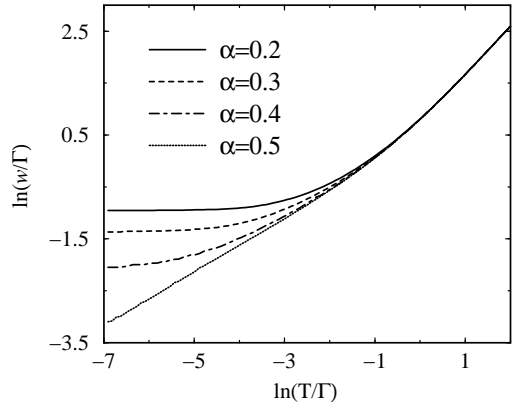}
\caption[]{\label{shirina} 
The width of the resonant conductance peak as a function of temperature for 
different values of the asymmetry parameter $\alpha$.}
\vspace*{-0.5cm}
\end{figure}

Another interesting issue is the shape of the resonance peak, especially its
width, $w(T)$, as a function of temperature, see Fig.\ref{shirina}. 
At high temperatures it decreases linearly upon lowering $T$ 
no matter how strong are the interactions. 
For $T\ll 1$, however, the correlation effects
become visible and the width $w(T)$ of the peak  
saturates at zero temperature unless the dot is symmetric. 
In the latter case $w(T)$ shrinks to zero with the 
exponent $1-g$ predicted in 
Ref.\cite{kanefisher}: $w(T) \sim T^{0.5}$.

It is in fact not difficult to evaluate the integral in 
Eq.(\ref{IVdefinition}) analytically. 
The general expression is complicated, so we shall
only present here some particular cases. 
To begin with we observe that there is an intimate relation between
our model at $\Delta=0$, $\alpha=1/2$ (resonant and symmetric 
case) and the $g=1/2$ solution for the conductance 
through a \emph{single} barrier ($G_s$) given in 
\cite{kanefisher}. Indeed, evaluating (\ref{1-D}) for the
case in question, we find (in linear response)
\begin{equation}                      \label{delta0}
G_{\Delta=0}(T)/G_0=\frac{1}{2\pi T}
\psi' \left( \frac{1}{2}+\frac{1}{2\pi T}\right)
\end{equation}
where $\psi$ is the $\psi$-function. Comparing with Ref.\cite{weiss}
(see also \cite{book}), we observe that $G_{\Delta=0}(T)/G_0=1-G_s(T)/G_0$
if $T$ in $G_s$ is measured in units of the backscattering strength. 
One can easily show that an analogous relation continue to
hold for the out--of--equilibrium current.

\begin{figure}
\vspace*{0.5cm}
\includegraphics[scale=0.38]{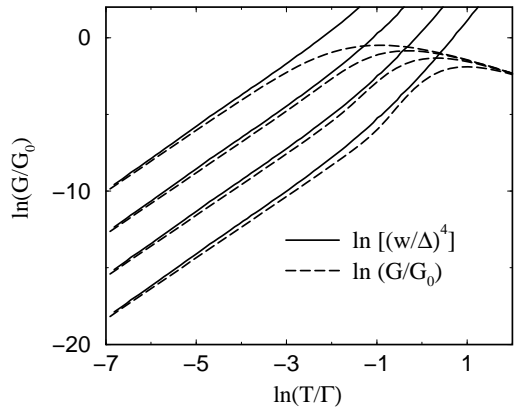}
\caption[]{\label{scaling2}
Comparison between the linear conductance (dashed lines) and the approximative
scaling function $(w/\Delta)^4$ (solid lines) for different values of
$\Delta$: from above $\Delta=0.5$, $1$, $2$, $4$. 
}
\vspace*{-0.5cm}
\end{figure}

Furthermore, for the linear conductance when $\alpha=1/2$ but
$\Delta\neq 0$ ($|\Delta|<1/2$) we obtain
\begin{eqnarray}
\label{deltanot0}
G_\Delta(T)/G_0&=&\frac{1}{2\pi T (\lambda_+^2-\lambda_-^2)}\left[
\lambda_+\psi' \left( \frac{1}{2}+\frac{\lambda_+}{2\pi T}\right)
\right. \nonumber \\ &-& \left.
\lambda_-\psi' \left( \frac{1}{2}+\frac{\lambda_-}{2\pi T}\right)
\right]\;,
\end{eqnarray}
where $\sqrt{2}\lambda_\pm=\sqrt{1-2\Delta^2\pm\sqrt{1-4\Delta^2}}$.
Same formula is valid in the resonant but asymmetric case
($\Delta=0$, $\alpha\neq 1/2$) if we substitute $\lambda_\pm \to
1/2\pm \sqrt{\alpha(1-\alpha)}$. 
This shows that the 
asymmetry parameter is equivalent to a small off-set of
the resonance (if the two act independently).
(Indeed, in terms of the Kondo analogy, the 
channel asymmetry is known to be relevant.)
For a strong off-set of the resonance ($|\Delta|>1/2$),
formula (\ref{deltanot0}) is not valid and should be replaced by:
\begin{equation}
\label{strong}
G_\Delta(T)/G_0=\frac{\Delta}{\pi T\sqrt{4\Delta^2-1}}
{\rm Im} \left[ e^{i\theta}
\psi' \left( \frac{1}{2}+\frac{\Delta e^{i\theta}}{2\pi T}\right)
\right]
\end{equation}
where $\theta=\tan^{-1}[\sqrt{4\Delta^2-1}/(1-2\Delta^2)]/2$. 

As pointed out in Ref.\cite{kanefisher}, for $\Delta, T\ll 1$,
the conductance should become a universal (scaling) function of
the ratio of the resonance width and the backscattering strength
(resonance off-set).
Indeed, taking the appropriate limit in formula (\ref{deltanot0}),
we obtain the exact scaling function at $g=1/2$:    
$G_\Delta(T)/G_0 =\tilde{G}_{g=1/2}(X)$, where
\begin{eqnarray}                  \label{scalingfunction}
\tilde{G}_{1/2}(X)= 1-\frac{2}{\pi^2}X^2
\psi'\left(\frac{1}{2}+\frac{2}{\pi^2}X^2\right) 
\end{eqnarray}
and the scaling variable is $X=\sqrt{\pi}\Delta/2T^{1/2}$
(that is $w(T)\simeq 2T^{1/2}/\sqrt{\pi}$ at small $T$).
We note that our scaling function $\tilde{G}_{g=1/2}(X)$
is by far more complicated than $\tilde{G}_{g\simeq 1}(X)=
1/(1+X^2)$ found by NG in weak coupling.
Furthermore we observe from our analytic expressions 
that beyond $\Delta, T \ll 1$ there is \emph{no} exact scaling.
However, upon determining $w(T)$ numerically and plotting $[w(T)/\Delta]^4$ 
versus the dimensionless conductance $G/G_0$, we obtain an 
\emph{approximate} numerical scaling as shown in Fig.\ref{scaling2}.
The same scaling function (\ref{scalingfunction}) holds in the resonant
($\Delta=0$) but weakly asymmetric case, when $\alpha-1/2$ is small. In that
situation $\alpha-1/2$ substitutes $\Delta$ in
the definition of $X$.

To summarise, we presented an explicit solution for the transport 
through a resonant level coupled to two LL leads. 
It turns out that for $g=1/2$ the Hamiltonian of the system 
can be mapped onto one similar to the two-channel Kondo
Hamiltonian in the Toulouse limit, solvable exactly. 
We obtained the full $I-V$ characteristics, which shows all the effects 
inherent to resonant tunnelling setups in LLs, including the scaling. 
Our solution 
confirms previous results obtained by means of the 
perturbation theory and goes beyond them. 
In future, it would be interesting to study deviations
from $g=1/2$ (in the spirit of Ref.\cite{weiss}) and to investigate
the effects of electron spin (and flavour). 

The authors would like to thank H. Grabert and R. Egger for a valuable 
discussion. This work was supported by the Landesstiftung Baden--W\"urttemberg
gGmbH (Germany), by the EPSRC of the UK under grants GR/N19359 and
GR/R70309, and by the EC network DIENOW.

\bibliography{restun}

\begin{thebibliography}{13}
\expandafter\ifx\csname natexlab\endcsname\relax\def\natexlab#1{#1}\fi
\expandafter\ifx\csname bibnamefont\endcsname\relax
  \def\bibnamefont#1{#1}\fi
\expandafter\ifx\csname bibfnamefont\endcsname\relax
  \def\bibfnamefont#1{#1}\fi
\expandafter\ifx\csname citenamefont\endcsname\relax
  \def\citenamefont#1{#1}\fi
\expandafter\ifx\csname url\endcsname\relax
  \def\url#1{\texttt{#1}}\fi
\expandafter\ifx\csname urlprefix\endcsname\relax\def\urlprefix{URL }\fi
\providecommand{\bibinfo}[2]{#2}
\providecommand{\eprint}[2][]{\url{#2}}

\bibitem[{\citenamefont{Dekker}(1999)}]{dekker}
\bibinfo{author}{\bibfnamefont{C.}~\bibnamefont{Dekker}},
  \bibinfo{journal}{Phys.~Today} \textbf{\bibinfo{volume}{52}},
  \bibinfo{pages}{No. 5,~22} (\bibinfo{year}{1999}), \bibinfo{note}{and
  references therein}.

\bibitem[{\citenamefont{Postma et~al.}(2001)\citenamefont{Postma, Teepen, Yao,
  Grifoni, and Dekker}}]{postma}
\bibinfo{author}{\bibfnamefont{H.~W.~C.} \bibnamefont{Postma}},
  \bibinfo{author}{\bibfnamefont{T.}~\bibnamefont{Teepen}},
  \bibinfo{author}{\bibfnamefont{Z.}~\bibnamefont{Yao}},
  \bibinfo{author}{\bibfnamefont{M.}~\bibnamefont{Grifoni}}, \bibnamefont{and}
  \bibinfo{author}{\bibfnamefont{C.}~\bibnamefont{Dekker}},
  \bibinfo{journal}{Science} \textbf{\bibinfo{volume}{293}},
  \bibinfo{pages}{76} (\bibinfo{year}{2001}).

\bibitem[{\citenamefont{Kane and Fisher}(1992)}]{kanefisher}
\bibinfo{author}{\bibfnamefont{C.~L.} \bibnamefont{Kane}} \bibnamefont{and}
  \bibinfo{author}{\bibfnamefont{M.~P.~A.} \bibnamefont{Fisher}},
  \bibinfo{journal}{Phys.~Rev.~B} \textbf{\bibinfo{volume}{46}},
  \bibinfo{pages}{15233} (\bibinfo{year}{1992}).

\bibitem[{\citenamefont{Nazarov and Glazman}(2002)}]{NG}
\bibinfo{author}{\bibfnamefont{Yu.~V.} \bibnamefont{Nazarov}} \bibnamefont{and}
  \bibinfo{author}{\bibfnamefont{L.~I.} \bibnamefont{Glazman}},
  \bibinfo{journal}{cond-mat/0209090}  (\bibinfo{year}{2002}).

\bibitem[{\citenamefont{Furusaki}(1998)}]{furusakidot}
\bibinfo{author}{\bibfnamefont{A.}~\bibnamefont{Furusaki}},
  \bibinfo{journal}{Phys. Rev. B} \textbf{\bibinfo{volume}{57}},
  \bibinfo{pages}{7141} (\bibinfo{year}{1998}).

\bibitem[{\citenamefont{Nayak et~al.}(1999)\citenamefont{Nayak, Fisher, Ludwig,
  and Lin}}]{nayak}
\bibinfo{author}{\bibfnamefont{C.}~\bibnamefont{Nayak}},
  \bibinfo{author}{\bibfnamefont{M.}~\bibnamefont{Fisher}},
  \bibinfo{author}{\bibfnamefont{A.}~\bibnamefont{Ludwig}}, \bibnamefont{and}
  \bibinfo{author}{\bibfnamefont{H.}~\bibnamefont{Lin}},
  \bibinfo{journal}{Phys. Rev. B} \textbf{\bibinfo{volume}{59}},
  \bibinfo{pages}{15694} (\bibinfo{year}{1999}).

\bibitem[{\citenamefont{Matveev}(1991)}]{matveevdot}
\bibinfo{author}{\bibfnamefont{K.~A.} \bibnamefont{Matveev}},
  \bibinfo{journal}{Sov. Phys. JETP} \textbf{\bibinfo{volume}{72}},
  \bibinfo{pages}{892} (\bibinfo{year}{1991}).

\bibitem[{\citenamefont{Gogolin et~al.}(1998)\citenamefont{Gogolin, Nersesyan,
  and Tsvelik}}]{book}
\bibinfo{author}{\bibfnamefont{A.~O.} \bibnamefont{Gogolin}},
  \bibinfo{author}{\bibfnamefont{A.~A.} \bibnamefont{Nersesyan}},
  \bibnamefont{and} \bibinfo{author}{\bibfnamefont{A.~M.}
  \bibnamefont{Tsvelik}}, \emph{\bibinfo{title}{Bosonization and Strongly
  Correlated Systems}} (\bibinfo{publisher}{Cambridge University Press},
  \bibinfo{year}{1998}).

\bibitem[{\citenamefont{Egger and Grabert}(1998)}]{eggergrabert}
\bibinfo{author}{\bibfnamefont{R.}~\bibnamefont{Egger}} \bibnamefont{and}
  \bibinfo{author}{\bibfnamefont{H.}~\bibnamefont{Grabert}},
  \bibinfo{journal}{Phys. Rev. B} \textbf{\bibinfo{volume}{58}},
  \bibinfo{pages}{10761} (\bibinfo{year}{1998}).

\bibitem[{\citenamefont{Emery and Kivelson}(1992)}]{emerykivelson}
\bibinfo{author}{\bibfnamefont{V.~J.} \bibnamefont{Emery}} \bibnamefont{and}
  \bibinfo{author}{\bibfnamefont{S.}~\bibnamefont{Kivelson}},
  \bibinfo{journal}{Phys.~Rev.~B} \textbf{\bibinfo{volume}{46}},
  \bibinfo{pages}{10812} (\bibinfo{year}{1992}).

\bibitem[{\citenamefont{Schiller and Hershfield}(1998)}]{SH}
\bibinfo{author}{\bibfnamefont{A.}~\bibnamefont{Schiller}} \bibnamefont{and}
  \bibinfo{author}{\bibfnamefont{S.}~\bibnamefont{Hershfield}},
  \bibinfo{journal}{Phys.~Rev.~B} \textbf{\bibinfo{volume}{58}},
  \bibinfo{pages}{14978} (\bibinfo{year}{1998}).

\bibitem[{\citenamefont{Datta}(1995)}]{datta}
\bibinfo{author}{\bibfnamefont{S.}~\bibnamefont{Datta}},
  \emph{\bibinfo{title}{Electronic Transport in Mesoscopic Systems}}
  (\bibinfo{publisher}{Cambridge University Press}, \bibinfo{year}{1995}).

\bibitem[{\citenamefont{Weiss et~al.}(1995)\citenamefont{Weiss, Egger, and
  Sassetti}}]{weiss}
\bibinfo{author}{\bibfnamefont{U.}~\bibnamefont{Weiss}},
  \bibinfo{author}{\bibfnamefont{R.}~\bibnamefont{Egger}}, \bibnamefont{and}
  \bibinfo{author}{\bibfnamefont{M.}~\bibnamefont{Sassetti}},
  \bibinfo{journal}{Phys. Rev. B} \textbf{\bibinfo{volume}{52}},
  \bibinfo{pages}{16707} (\bibinfo{year}{1995}).

\end{thebibliography}
\end{document}